\documentclass[aps,superscriptaddress,twocolumn,pre,longbibliography,floatfix]{revtex4-1}

\usepackage{amsmath,amsthm,amsfonts,amssymb,bm,graphicx,color,times}
\usepackage[colorlinks={true}, citecolor={blue}, filecolor={blue}, linkcolor={blue}, urlcolor={blue}]{hyperref}
\usepackage[caption=false]{subfig}

\usepackage{soul}

\newcommand{\rmd}{\mathrm{d}}
\allowdisplaybreaks

\begin{document}

\title{Thermodynamic consistency of the optomechanical master equation}

\author{Muhammad T.\ Naseem}
\affiliation{Department of Physics, Ko\c{c} University, Sar{\i}yer, \.Istanbul, 34450, Turkey}
\author{Andr\'e Xuereb}
\affiliation{Department of Physics, University of Malta, Msida MSD 2080, Malta}
\author{\"{O}zg\"{u}r E.\ M\"{u}stecapl{\i}o\u{g}lu}
\email[Electronic address:\ ]{omustecap@ku.edu.tr}
\affiliation{Department of Physics, Ko\c{c} University, Sar{\i}yer, \.Istanbul, 34450, Turkey}

\begin{abstract}
We investigate the thermodynamic consistency of the master equation description of heat transport through an optomechanical system attached to two heat baths, one optical and one mechanical. We employ three different master equations to describe this scenario:\ (i)~The standard master equation used in optomechanics, where each bath acts only on the resonator that it is physically connected to; (ii)~the so-called dressed-state master equation, where the mechanical bath acts on the global system; and (iii)~what we call the global master equation, where both baths are treated non-locally and affect both the optical and mechanical subsystems. Our main contribution is to demonstrate that, under certain conditions including when the optomechanical coupling strength is weak, the second law of thermodynamics is violated by the first two of these pictures. In order to have a thermodynamically consistent description of an optomechanical system, therefore, one has to employ a global description of the effect of the baths on the system.
\end{abstract}
\maketitle

\section{Introduction}\label{sec:intro}
The field of optomechanics~\cite{review1, review2} investigates composite systems where an optical resonator is coupled to a mechanical oscillator. A significant portion of the studies in the field focus on its promise for testing fundamental quantum laws using macroscopic objects, constructing probes for tiny forces with quantum-limited sensitivity, generating non-classical states, and as interfaces for applications in hybrid quantum information systems~\cite{ap1, ap2, ap3, ap4, ap5, ap6, ap7, ap8, ap9, ap10, ap11, ap12, ap13, ap14, ap15, ap16, ap17, ap18}. Attention has, more recently, been devoted to the thermodynamic applications of optomechanical systems~\cite{Zhang_OM_HE, Zhang_OM_HE14, Phonon_cooling, Sidemodes_master, Dong_HE2015, Brunelli_OutEqm, Mari_HE, Bathee_HE2016, Zhang_HE2017}, including proposals for optomechanical quantum heat engines~\cite{qhe_scr} and heat transport through optomechanical arrays~\cite{heat_transport_arrays}. Despite all this research, and somewhat surprisingly, it appears that a thermodynamically consistent open system description of the optomechanical interaction that is valid at arbitrary coupling strength is still lacking.

In part due to the limitations of current experimental setups, the typical description of an optomechanical system is restricted in validity to the weak coupling regime. The dynamical behavior in this scenario is typically studied using what we will refer to as the standard master equation (SME), where the heat baths connected to the system are assumed to influence only the system that they are attached to~\cite{breuer2002}. As the coupling strength grows and the system enters the strong-coupling regime, it has been suggested~\cite{Dress_master} to use the so-called dressed-state master equation (DSME). In effect, this description includes the influence of the mechanical heat bath on the optical resonator, but neglects the effect of the optical heat bath on the mechanical oscillator; one can say that the optical reservoir is local whereas the mechanical one global~\cite{kosloff_local_violate}. Recent proposals~\cite{ap19,ap20} have suggested ways in which this regime may be rendered accessible, highlighting the need for understanding which description of the dynamics is to be used. Indeed, consistency of these different master equations with the laws of thermodynamics is not guaranteed; it is known, for example, that coupled simple harmonic resonators in certain parameter regimes require a fully global treatment of the reservoirs to ensure thermodynamic consistency~\cite{ap21, kosloff_local_violate, violation-Stockburger, Patrick-LocalVsGlobal}.

Our main objective in this paper is to systematically examine the heat transport through an optomechanical system from the point of view the first and second laws of thermodynamics. We highlight the failure of both the aforementioned master equations to enforce consistency with the second law of thermodynamics, and propose a method based upon a global master equation (GME) to ensure consistency at arbitrary optomechanical coupling strength. In particular we provide evidence showing that consistency with the second law of thermodynamics requires phonon sideband modes to be included in the master equation when the temperature of the mechanical bath is greater than that of the optical bath.

This paper is organized as follows. In Sec.~\ref{sec:Model}, we outline our basic model and present the three master equations that are the subject of our study. In Sec.~\ref{sec:results} we compare the three approaches in terms of their consistency with the second law of thermodynamics. We then sum up briefly and give our conclusions in Sec.~\ref{sec:conclusion}.

\section{The Model}
\label{sec:Model}
\begin{figure}[t!]
  \centering
  \includegraphics[scale=0.5]{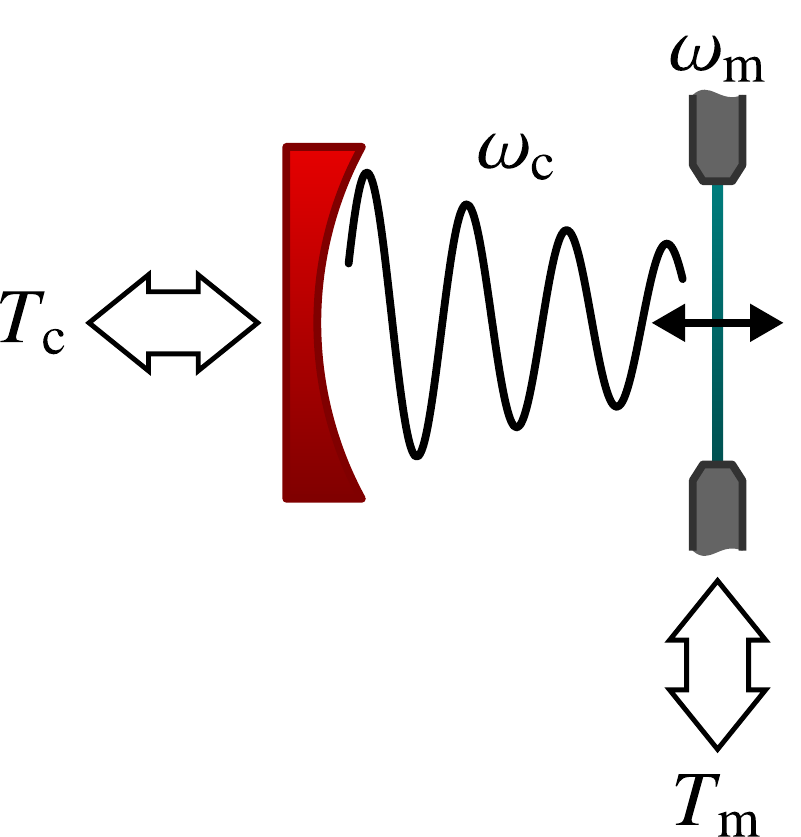}
  \caption{Schematic illustration of an optomechanical system, consisting of a mechanical resonator with frequency $\omega_\text{m}$ coupled to an optical resonator with frequency $\omega_\text{c}$. In this illustration, an optical cavity has a mobile end mirror (shown on the right) whose motion is assumed to be harmonic. The two resonators are coupled to heat baths, at temperatures $T_\text{m}$ and $T_\text{c}$, respectively. We assume that these two baths are independent and can posses any finite non-negative temperature.}
  \label{fig:Setup}
\end{figure}

Our model consists of a Fabry--P\'erot cavity one of whose end mirrors is allowed to move; this model is shown schematically in Fig.~\ref{fig:Setup} and is representative of a large class of physically equivalent systems containing a localized electromagnetic field mode interacting with a mechanical oscillator. The optical (mechanical) resonator in our model has frequency $\omega_\text{c}$ ($\omega_\text{m}$) and is attached to a thermal bath at temperature $T_\text{c}$ ($T_\text{m}$). Both these baths are independent and can possess any non-negative finite temperature. Since the description of the baths depends on the specific approach followed, as detailed below, we defer this discussion to the forthcoming subsections.

The Hamiltonian governing the evolution of the isolated system consisting of the optical and mechanical modes is (we use units in our Hamiltonians such that $\hbar=1$ for convenience)
\begin{equation}
\hat{H}=\omega_\text{c}\hat{a}^{\dagger}\hat{a}+\omega_\text{m}\hat{b}^{\dagger}b-g\hat{a}^{\dagger}\hat{a}(\hat{b}+\hat{b}^{\dagger}).
\end{equation}
The first two terms in $\hat{H}$ are energies of the optical and mechanical modes, respectively, whereas third term denotes the optomechanical interaction with single-photon coupling strength $g$. We denote the annihilation (creation) operator of the optical mode by $\hat{a}$ ($\hat{a}^{\dagger}$) and of the mechanical mode by $\hat{b}$ ($\hat{b}^{\dagger}$).

\subsection{Three different master equations}
We will now proceed to add to $\hat{H}$ the interaction of the two isolated modes with two independent thermal baths. In this subsection we shall consider three different master equations which may be used to describe the dynamics of the reduced density matrix of the system after the two baths have been traced out.

\subsubsection{The standard master equation (SME)}
Let us first consider the case where each of the two degrees of freedom is coupled to an independent bath mode; these baths can therefore be considered local, in the sense that they interact with the localized field operators $\hat{a}$ and $\hat{b}$. The Hamiltonian describing the full system is
\begin{multline}
\hat{H}_\text{tot}=\hat{H}+\sum_\lambda\Bigl[\omega_{\text{c},\lambda}\hat{c}^{\dagger}_{\text{c},\lambda}\hat{c}_{\text{c},\lambda}+g_{\text{c},\lambda}(\hat{a}^{\dagger}\hat{c}_{\text{c},\lambda}+\hat{a}\hat{c}^{\dagger}_{\text{c},\lambda})\\
+\omega_{\text{m},\lambda}\hat{c}^{\dagger}_{\text{m},\lambda}\hat{c}_{\text{m},\lambda}+g_{\text{m},\lambda}(\hat{b}^{\dagger}\hat{c}_{\text{m},\lambda}+\hat{b}\hat{c}^{\dagger}_{\text{m},\lambda})\Bigr],
\end{multline}
The sum in $\hat{H}_\text{tot}$ runs over the infinite number of bath modes, indexed by $\lambda$ (which may be regarded as a continuous or discrete index) for both the optical and mechanical baths. The first term in the sum represents the free Hamiltonians of the optical bath modes, $\omega_{\text{c},\lambda}$ being the frequency of the bath mode indexed by $\lambda$ and $\hat{c}_\text{c}$ ($\hat{c}_\text{c}^{\dagger}$) its annihilation (creation) operator. The second term represents the interaction between these bath modes and the optical resonator, where the interaction with the bath mode indexed by $\lambda$ is governed by a strength $g_{\text{c},\lambda}$. The last two terms in the sum are analogous to these first two, but describe the mechanical bath modes and their interaction with the mechanical resonator. The optical and mechanical baths are assumed to be at thermal equilibrium at temperatures $T_\text{c}$ and $T_\text{m}$, respectively.

The standard way of deriving the master equation starts off by making the Born--Markov approximation, the details of which and whose regime of validity can be found in Refs.~\cite{breuer2002, critcal-MME}. In a second step, the weak coupling approximation is made, which finally results in a local master equation without the need to make any secular approximation, i.e.,
\begin{equation}\label{eq:SME}
\frac{\rmd\hat{\rho}}{\rmd t}=-i[\hat{H},\hat{\rho}]+\mathcal{L}^\text{(s)}_\text{c} \hat{\rho}+\mathcal{L}^\text{(s)}_\text{m} \hat{\rho},
\end{equation}
where
\begin{align}
\mathcal{L}^\text{(s)}_\text{c} \hat{\rho}&=G_\text{c}(\omega_\text{c})D[\hat{a}]\hat{\rho}+G_\text{c}(-\omega_\text{c})D[\hat{a}^{\dagger}]\hat{\rho}\ \text{and}\\
\mathcal{L}^\text{(s)}_\text{m} \hat{\rho}&=G_\text{m}(\omega_\text{m})D[\hat{b}]\hat{\rho}+G_\text{m}(-\omega_\text{m})D[\hat{b}^{\dagger}]\hat{\rho}
\end{align}
are the Liouville super-operators of the optical and mechanical baths, respectively. In these equations
\begin{equation}
D[\hat{o}]\hat{\rho}:=\tfrac12(2\hat{o}\hat{\rho}\hat{o}^{\dagger}-\hat{o}^{\dagger}\hat{o}\hat{\rho}-\hat{\rho}\hat{o}^{\dagger}\hat{o})
\end{equation}
is the Linblad dissipator. The spectral density functions $G_x(\omega)$ ($x=\text{c},\text{m}$) of the thermal baths are given by
\begin{align}
G_x(\omega)&=\gamma_x(\omega)[1 + \bar{n}_x(\omega)]\ \text{and}\\
G_x(-\omega)&=\gamma_x(\omega)\bar{n}_x(\omega),
\end{align}
where
\begin{equation}
\bar{n}_x(\omega)=\frac{1}{e^{\hbar\omega/(k_\text{B}T_x)}-1}
\end{equation}
are the Bose--Einstein distributions of the excitations in the baths, with $k_\text{B}$ being the Boltzmann constant. The coefficients
\begin{equation}
\gamma_x(\omega)=2\pi\hbar\frac{f_x(\omega)g_x(\omega)^2}{\omega}
\end{equation}
are determined by the densities of modes of the baths, $f_x(\omega)$, and the interaction strengths between the baths and their corresponding resonators, $g_x(\omega)$. In the following, we assume strictly Ohmic baths with flat densities of modes, in which case the $\gamma_x(\omega)$ become independent of $\omega$ and can be denoted by $\kappa_x:=\gamma_x(\omega)$.

\subsubsection{The dressed-state master equation (DSME)}
In contrast with the standard master equation, where each bath couples to a local degree of freedom, this is a semi-global approach. To derive the DSME, the reduced system Hamiltonian is diagonalized by means of a polaron transformation, and the system--bath interaction is described in this new basis. Next, one makes the usual Born--Markov approximation. An assumption is then made whereby the bath attached to the mechanical resonator couples to both degrees of freedom, but where the other bath couples only to the optical resonator. This approximation is valid for sufficiently flat spectral density and $\omega_\text{c}\gg\omega_\text{m}$, in which case phonon side-bands can be ignored. In the DSME one further assumes that $\bar{n}_\text{m}\gg 1$. A detailed derivation of DSME is presented in Ref.~\cite{Dress_master} and yields, finally,
\begin{equation}\label{eq:DSME}
\frac{\rmd\hat{\rho}}{\rmd t}=-i[\hat{H},\hat{\rho}] +\mathcal{L}^\text{(d)}_\text{c}\hat{\rho} + \mathcal{L}^\text{(d)}_\text{m}\hat{\rho}+\mathcal{L}^\text{(d)}_{\text{m},\text{d}}\hat{\rho},
\end{equation}
where
\begin{align}
\mathcal{L}^\text{(d)}_\text{c} \hat{\rho}&=G_\text{c}(\omega_\text{c})D[\hat{a}]\hat{\rho}+G_\text{c}(-\omega_\text{c})D[\hat{a}^{\dagger}]\hat{\rho},\\
\mathcal{L}^\text{(d)}_\text{m} \hat{\rho}&=G_\text{m}(\omega_\text{m})D[\hat{b}-\alpha\hat{n}_\text{c}]\hat{\rho}\nonumber\\
&\qquad+G_\text{m}(-\omega_\text{m})D[\hat{b}^{\dagger}-\alpha\hat{n}_\text{c}]\hat{\rho},\ \text{and}\\
\mathcal{L}^\text{(d)}_{\text{m},\text{d}} \hat{\rho}&=4(\kappa_\text{m} T_\text{m}/\omega_\text{m}) \alpha^2D[\hat{n}_\text{c}]\hat{\rho}.
\end{align}
The first of these two equations describe the dissipation of the optical and mechanical mode, respectively, and the last equation represents dephasing of the optical mode. Furthermore,  $\hat{n}_\text{c}=\hat{a}^\dagger\hat{a}$ and $\alpha=g/\omega_\text{m}$. We note that this master equation reduces identically to the SME in the limit $g\to0$.

\subsubsection{The global master equation (GME)}
The derivation in this case is similar to previous case, except that here both baths are treated on an equal (global) footing, and that phonon side-bands are not ignored. In the interaction picture the coupling between the baths and oscillators is characterized by the interaction Hamiltonians
\begin{subequations}\label{eq:sysbath}
\begin{align}
\hat{H}_\text{c,B}&=\hat{a}^{\dagger}(t)\hat{c}_\text{c}(t)+\hat{a}(t)\hat{c}^{\dagger}_\text{c}(t)\ \text{and}\\
\hat{H}_\text{m,B}&=\hat{b}^{\dagger}(t)\hat{c}_\text{m}(t)+\hat{b}(t)\hat{c}^{\dagger}_\text{m}(t).
\end{align}
\end{subequations}
Here we have defined $\hat{o}(t)=e^{i\hat{H}t}\hat{o}e^{-i\hat{H}t}$ and $\hat{c}_\text{c}$ ($\hat{c}_\text{m}$) are non-normalized optical (mechanical) bath operators, with $\hat{c}_\text{c}(t)=\sum_{\lambda}g_{\text{c},\lambda}e^{-i\omega_{\text{c},\lambda}t}\hat{c}_{\text{c},\lambda}$ and $\hat{c}_\text{m}(t)=\sum_{\lambda}g_{\text{m},\lambda}e^{-i\omega_{\text{m},\lambda}t}\hat{c}_{\text{m},\lambda}$.

The Hamiltonian of the reduced system, $\hat{H}$, can be diagonalized using the transformation
\begin{equation}
\hat{S} = e^{-\alpha\hat{a}^{\dagger}\hat{a}(\hat{b}^{\dagger}-\hat{b})},
\end{equation}
following which the Hamiltonian takes the form
\begin{equation}
\tilde{H}=\omega_\text{c}\tilde{a}^{\dagger}\tilde{a}+\omega_\text{m}\tilde{b}^{\dagger}\tilde{b}-\frac{g^2}{\omega_\text{m}}(\tilde{a}^{\dagger}\tilde{a})^2.
\end{equation}
The transformed operators then read
\begin{align}
\tilde{a}&=\hat{a} e^{-\alpha(\hat{b}^{\dagger}-\hat{b})}\ \text{and} \\
\tilde{b}&=\hat{b} - \alpha \hat{a}^{\dagger}\hat{a}.
\end{align}
The system operators in Eqs.~(\ref{eq:sysbath}) evaluate to
\begin{subequations}\label{eq:interaction}
\begin{align}
\hat{a}(t)&=\tilde{a}e^{-i\omega_\text{c}t} \sum_{n=0}^{\infty}\alpha^{n}(\tilde{b}e^{-i\omega_\text{m}t}-\tilde{b}^{\dagger}e^{i\omega_\text{m}t})^{n}\ \text{and}\\
\hat{b}(t)&=\tilde{b}e^{-i\omega_\text{m}t}+\alpha\tilde{a}^{\dagger}\tilde{a}.
\end{align}
\end{subequations}
From Eqs.~(\ref{eq:interaction}) the master equation can be derived by making standard Born--Markov and secular approximations. For simplicity we consider four side-bands, resulting in the master equation
\begin{align}\label{eq:GME}
\frac{\rmd\tilde{\rho}}{\rmd t}=\ &\mathcal{L}^\text{(g)}_\text{c} \tilde{\rho}+\mathcal{L}^\text{(g)}_\text{m} \tilde{\rho}+\mathcal{L}^\text{(g)}_{\text{m},\text{d}} \tilde{\rho},
\end{align}
where the dissipative and dephasing terms are given by
\begin{align}
\mathcal{L}^\text{(g)}_\text{c} \tilde{\rho}=\ &G_\text{c}(\omega_\text{c})\bigl\{D[\tilde{a}]+\alpha^{2}\bigl(D[\tilde{a}\tilde{b}\tilde{b}^{\dagger}]+D[\tilde{a}\tilde{b}^{\dagger}\tilde{b}]\bigr)\bigr\}\tilde{\rho}\nonumber\\
&+G_\text{c}(-\omega_\text{c})\bigl\{D[\tilde{a}^{\dagger}]+\alpha^{2}\bigl(D[\tilde{a}^{\dagger}\tilde{b}\tilde{b}^{\dagger}]\nonumber\\
&\qquad\qquad+D[\tilde{a}^{\dagger}\tilde{b}^{\dagger}\tilde{b}]\bigr)\bigr\}\tilde{\rho}\nonumber\\
&+\sum_{n=1,2}\Bigl\{\alpha^{n} G_\text{c}(\omega_\text{c} + n\omega_\text{m})D[\tilde{a}\tilde{b}^{n}]\tilde{\rho}\nonumber\\
&\qquad\qquad+\alpha^{n} G_\text{c}(-\omega_\text{c} - n\omega_\text{m})D[\tilde{a}^{\dagger}\tilde{b}^{\dagger n}]\tilde{\rho}\nonumber\\
&\qquad\qquad+\alpha^{n} G_\text{c}(\omega_\text{c} - n\omega_\text{m})D[\tilde{a}\tilde{b}^{\dagger n}]\tilde{\rho}\nonumber\\
&\qquad\qquad+\alpha^{n} G_\text{c}(-\omega_\text{c} + n\omega_\text{m})D[\tilde{a}^{\dagger}\tilde{b}^{n}]\tilde{\rho}\Bigr\},\\
\mathcal{L}^\text{(g)}_\text{m} \tilde{\rho}=&\ G_\text{m}(\omega_\text{m})D[\tilde{b}-\alpha\hat{n}_\text{c}]\tilde{\rho}\nonumber\\
&+G_\text{m}(-\omega_\text{m})D[\tilde{b}^{\dagger}-\alpha\hat{n}_\text{c}]\tilde{\rho},\ \text{and}\\
\mathcal{L}^\text{(g)}_{\text{m},\text{d}} \tilde{\rho}=&\ G_\text{m}(0)\alpha^2D[\hat{n}_\text{c}]\tilde{\rho}.
\end{align}
In principle all the phonon side-bands should be considered but, as we will show, for consistency with the second law of thermodynamics it is sufficient to consider first few side-bands even for rather strong single-photon coupling. In this work we will consider up to eight phonon side-bands and refer to Eq.~(\ref{eq:GME}) as global master equation with two and four side-bands as GME2 and GME4, respectively. The dissipators of for six and eight side-bands, giving rise to GME6 and GME8, respectively, are too cumbersome to report here. We note that, as required, in the limit $g\rightarrow 0$, Eq.~(\ref{eq:GME}) reduces to SME~(\ref{eq:SME}).

\subsection{Entropy production rate and heat current}
According to the first law of thermodynamics, the energy of an isolated system is conserved and can be split into heat and work~\cite{first_law}. For a quantum system, the dynamical version of second law of thermodynamics states that the entropy production rate of an isolated system remains non negative~\cite{second_law_kosl}:
\begin{equation}\label{eq:secondlaw}
\xi:=\frac{\rmd S}{\rmd t}-\sum_x\frac{\hbar\mathcal{J}_x}{k_\text{B}T_x}\geq0.
\end{equation}
In this equation $S$ is the von Neumann entropy, given by $S(\hat{\rho})=-\text{Tr}(\hat{\rho}\log\hat{\rho})$; in second term $\mathcal{J}_x$ represents the heat flux from the bath, which is given as~\cite{second_law_kosl}
\begin{equation}\label{eq:heatcurrent}
\mathcal{J}_x=\text{Tr}\{(\mathcal{L}^{(\gamma)}_x \hat{\rho})\hat{H}\},
\end{equation}
where $\mathcal{L}^{(\gamma)}_x$ represents the dissipative terms for the SME ($\gamma=\text{s}$), the DSME ($\gamma=\text{d}$), or the GME ($\gamma=\text{g}$), as the case requires. Noting that $\text{Tr}\{(\mathcal{L}^{(\gamma)}_{\text{m},\text{d}} \hat{\rho})\hat{H}\}=0$ ($x=\text{d},\text{g}$), at steady state we therefore have $\mathcal{J}_\text{c}+\mathcal{J}_\text{m}=0$ for all three models, which corresponds to the energy balance dictated by the first law of thermodynamics~\cite{second_law_kosl}.

In App.~\ref{app:heatflow} we solve the SME and DSME models thus yielding the steady-state entropy production rate,
\begin{equation}
\xi^\text{ss}=\frac{\hbar g\kappa_\text{c}}{k_\text{B}}\biggl(\frac{2g\omega_\text{m}+\alpha\kappa_\text{m}\kappa}{\omega_\text{m}^2+\kappa^2}\biggr)\bar{n}_\text{c}(\bar{n}_\text{c}+1)\biggl(\frac{1}{T_\text{m}}-\frac{1}{T_\text{c}}\biggr),
\end{equation}
where $\kappa=\kappa_\text{c}+\tfrac{\kappa_\text{m}}{2}$ and $\bar{n}_\text{c}:=\bar{n}_\text{c}(\omega_\text{c})$. All the factors in this expression but the last are non-negative, such that the sign of $\xi^\text{ss}$ is dictated exclusively by the relative magnitude of the non-negative temperatures $T_\text{c}$ and $T_\text{m}$. We are not aware of any concise expression for the steady-state entropy production rate predicted by the GME.

\begin{figure}
  \centering
  \subfloat[Steady state heat currents.]{\includegraphics[width=0.5\textwidth]{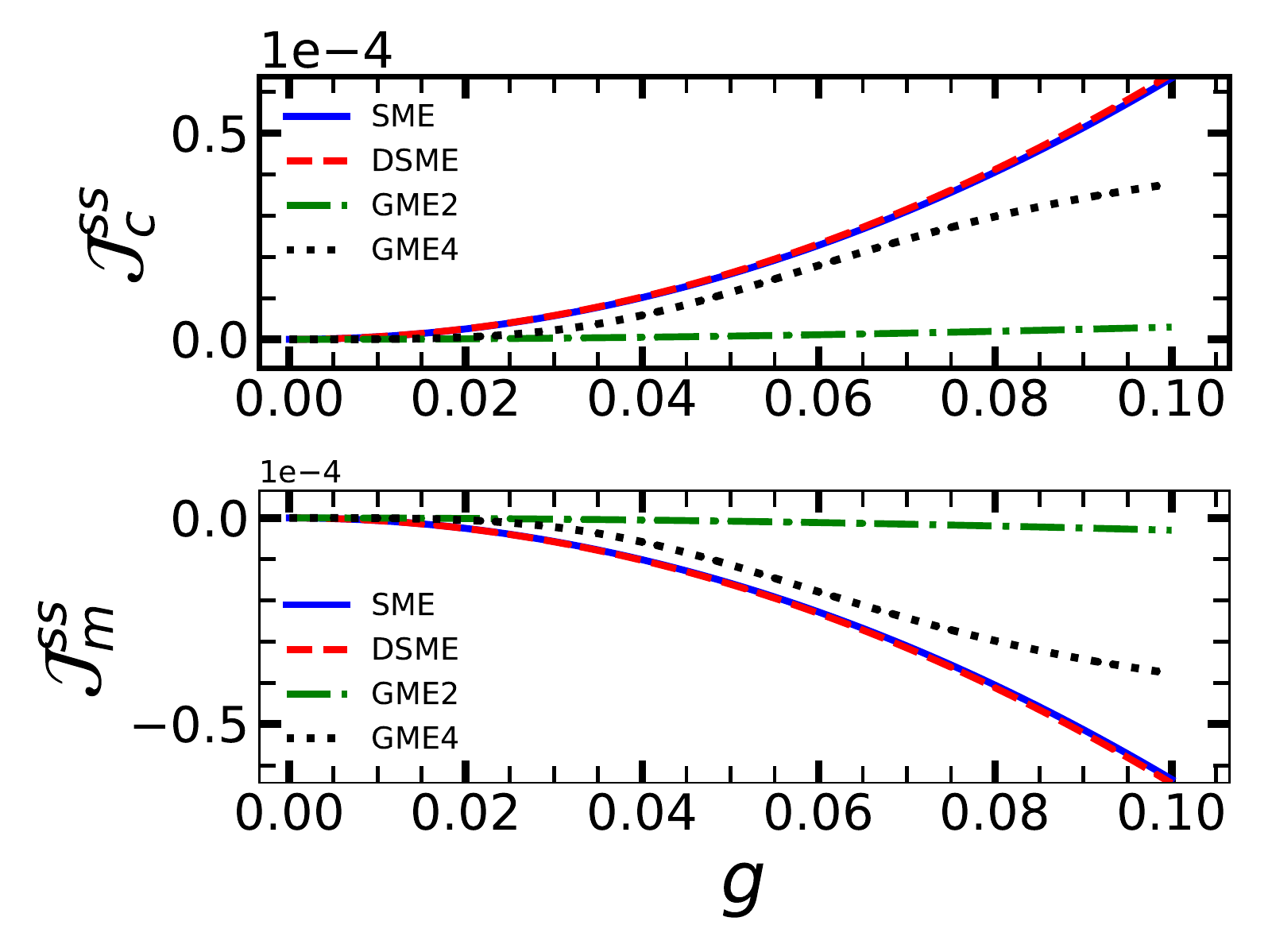}\label{fig:TmltTca}}\\
  \subfloat[Rates of entropy production.]{\includegraphics[width=0.5\textwidth]{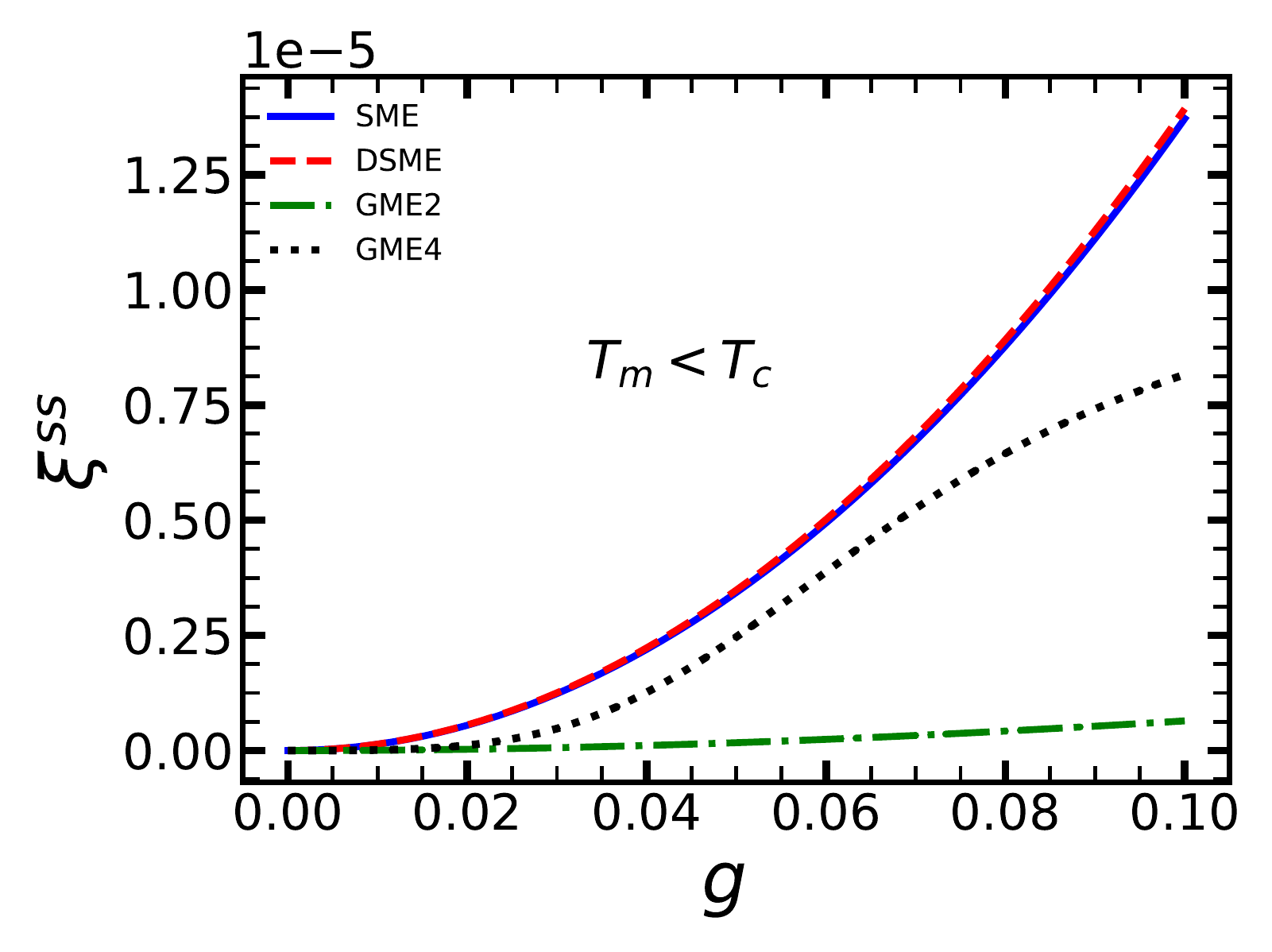}\label{fig:TmltTcb}}
  \caption{(a)~The steady state heat currents $\mathcal{J}^\text{ss}_\text{c}$ and $\mathcal{J}^\text{ss}_\text{m}$, and (b)~the rates of entropy production $\xi^\text{ss}$, as a function of the normalized coupling strength $g$ for $T_\text{c} > T_\text{m}$. The results from the standard master equation (SME) are illustrated by the solid blue curves, those for the dressed-state master equation (DSME) by the red dashed curves, and those for the global master equation (GME) by the green dash--dotted curves (two side-bands) and the black dotted curves (four side-bands). Parameters: $\kappa_\text{c}=0.02$, $\kappa_\text{m}=0.005$, $\omega_\text{c}=1, \omega_\text{m}=0.06$, $T_\text{c}=106$\,mK, and $T_\text{m}=101$\,mK.}
  \label{fig:TmltTc}
\end{figure}

\section{Results}\label{sec:results}
In this section we shall make use of the three different master equations developed in the previous section to present a comparative analysis of their consistency with the first laws of thermodynamics for selected parameters. In our simulations we use the Python quantum toolbox QuTiP~\cite{qutip} to solve he
master equations, and we take parameters relevant to circuit QED optomechanical simulators~\cite{Jhonson_OM-like}:\ $\omega_\text{c}=2\pi\times10$\,GHz, as well as $\omega_\text{m}=2\pi\times600$\,MHz, $\kappa_\text{c}=2\pi\times200$\,MHz, and $\kappa_\text{m}=2\pi\times50$\,MHz. From this point on, all our frequencies will be rescaled by $\omega_\text{c}$ and thus rendered dimensionless.

Recall that we are considering the situation when the optical and mechanical resonators are coupled to two  distinct thermal baths at temperatures $T_\text{c}$ and $T_\text{m}$, respectively. The two baths are independent and can posses any finite non-negative temperature. We shall analyze our models in three different cases: $T_\text{c} > T_\text{m}$, $T_\text{c} = T_\text{m}$, and $T_\text{c} < T_\text{m}$.

\subsection{Mechanics colder than optics, $T_\mathrm{c} > T_\mathrm{m}$}
Since the optical and mechanical oscillators are connected to two distinct heat baths, two heat currents, $\mathcal{J}_\text{c}$ and $\mathcal{J}_\text{m}$, are present in the system. The heat current equations for the SME and DSME are given in App.~\ref{app:heatflow}, while for the GME we calculate heat currents numerically. Figure~\ref{fig:TmltTca} shows the steady-state heat currents $\mathcal{J}^\text{ss}_\text{c}$ and $\mathcal{J}^\text{ss}_\text{m}$ as a function of the coupling strength $g$. When the two sub-systems are uncoupled, both heat currents are zero as expected. The local (SME, DSME) and global (GME) approaches coincide in the limit $g\rightarrow 0$. This result is in contrast with that in Ref.~\cite{Patrick-LocalVsGlobal}, in which a comparison between local and global master equations is performed for two interacting harmonic oscillators with coupling $(\hat{a}^{\dagger}\hat{b} + \hat{a}\hat{b}^{\dagger})$. In their case, when the two harmonic oscillators are uncoupled and have the same frequency, the SME gives correct (zero) heat currents but the GME yields unphysical (non-zero) heat currents. The divergence of SME and GME in Ref.~\cite{Patrick-LocalVsGlobal} for the limit $g\rightarrow 0$ is due to failure of secular approximation in this regime. In contrast, in our case the GME reduces identically to the SME in this limit, and we recover consistency with the second law of thermodynamics. However if $\omega_\text{c}=\omega_\text{m}$ and the optomechanical coupling strength $g<\kappa_x$ ($x=\text{c},\text{m}$) is not very small, then the secular approximation is not well justified and the GME will fail to yield consistent results~\cite{critcal-MME, Patrick-LocalVsGlobal}. Since we are considering an optomechanical system in which this parameter regime is not accessible, we are justified in using the GME for our dynamical description.

The heat currents $\mathcal{J}^\text{ss}_\text{c}$ and $\mathcal{J}^\text{ss}_\text{m}$ increase as the coupling strength grows. The heat current flows from the hot bath to the cold one, i.e., $\mathcal{J}^{\text{ss}}_\text{c}$ is positive and $\mathcal{J}^{\text{ss}}_\text{m}$ is negative. Moreover, both currents are equal at steady state, satisfying the energy conservation requirement. The inclusion of up to four phonon side-bands does not effect the qualitative behavior of the heat currents in this case, but it results in a change of magnitude as compared to the SME and the DSME.

Fig.~\ref{fig:TmltTcb} shows the entropy production rate, which remains non-negative for all the dynamical equations considered, even in strong coupling regime. Therefore, in the case of $T_\text{c}>T_\text{m}$ all three dynamical equations are thermodynamically consistent for both weak and strong coupling regimes, although they do predict different dynamical behaviors, especially when $g\gtrsim\omega_\text{m}$.

\begin{figure}
  \centering
  \includegraphics[width=0.5\textwidth]{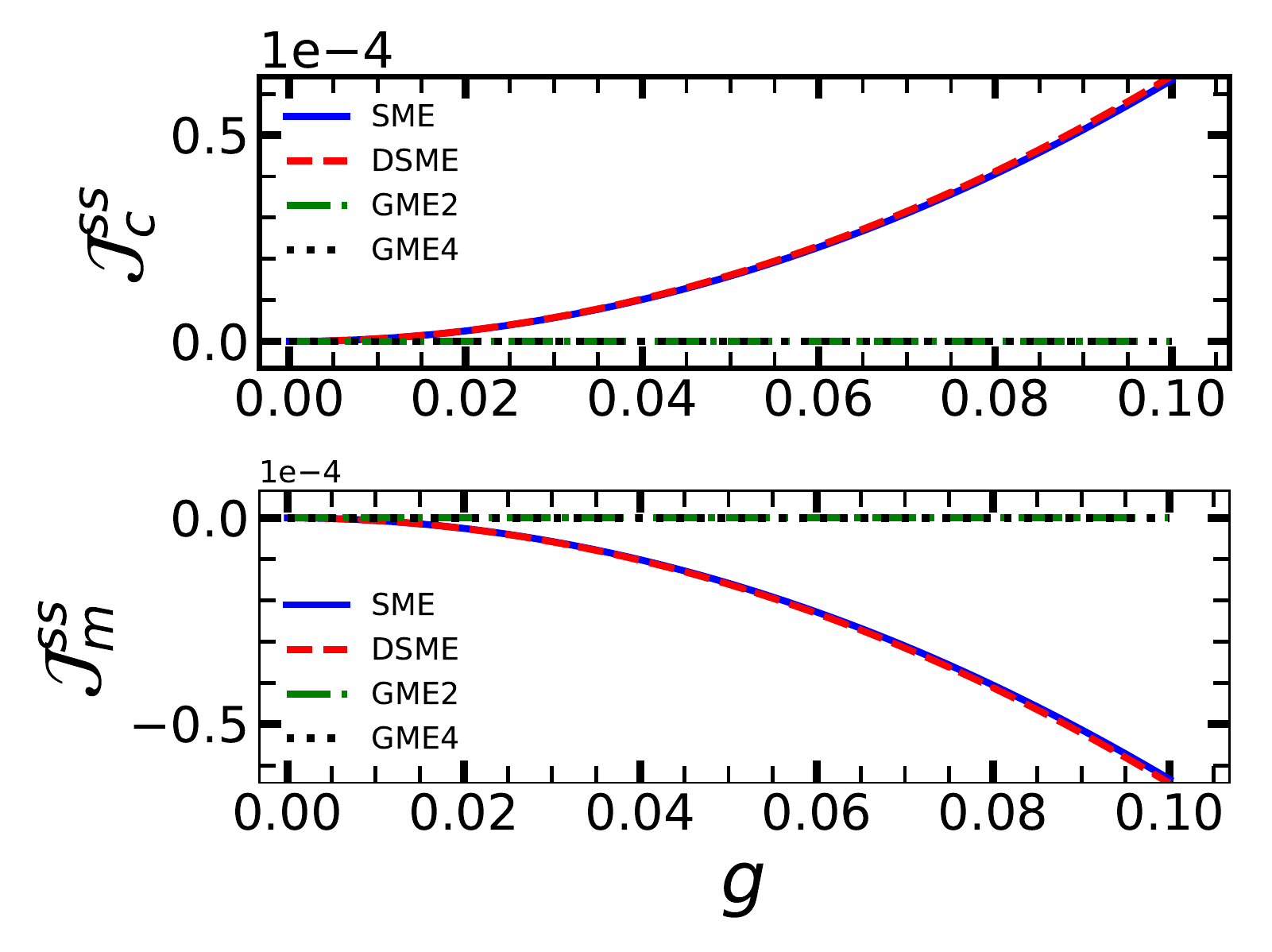}
  \caption{The steady state heat currents $\mathcal{J}^\text{ss}_\text{c}$ and $\mathcal{J}^\text{ss}_\text{m}$ as a function of normalized coupling strength $g$ for $T_\text{c} = T_\text{m}$. The results from the standard master equation (SME) are illustrated by the solid blue curves, those for the dressed-state master equation (DSME) by the red dashed curves, and those for the global master equation (GME) by the green dash--dotted curves (two side-bands) and the black dotted curves (four side-bands). Parameters: $T_\text{c}=T_\text{m}=106$\,mK, with the rest of the parameters as in Fig.\ \ref{fig:TmltTc}.}
  \label{fig:TmeqTc}
\end{figure}

\begin{figure}
  \centering
  \subfloat[Steady state heat currents.]{\includegraphics[width=0.5\textwidth]{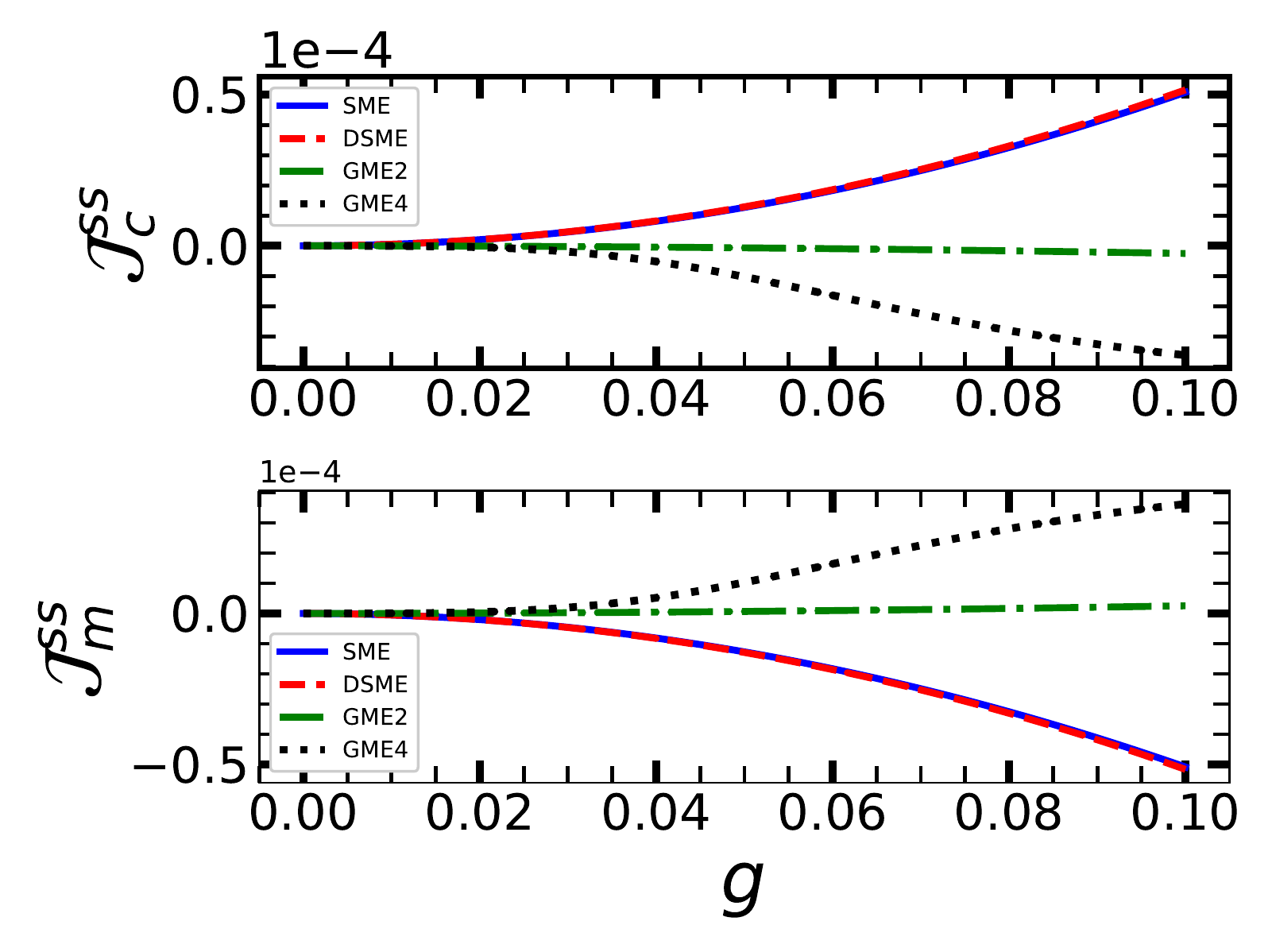}\label{fig:TmgtTca}}\\
  \subfloat[Rates of entropy production.]{\includegraphics[width=0.5\textwidth]{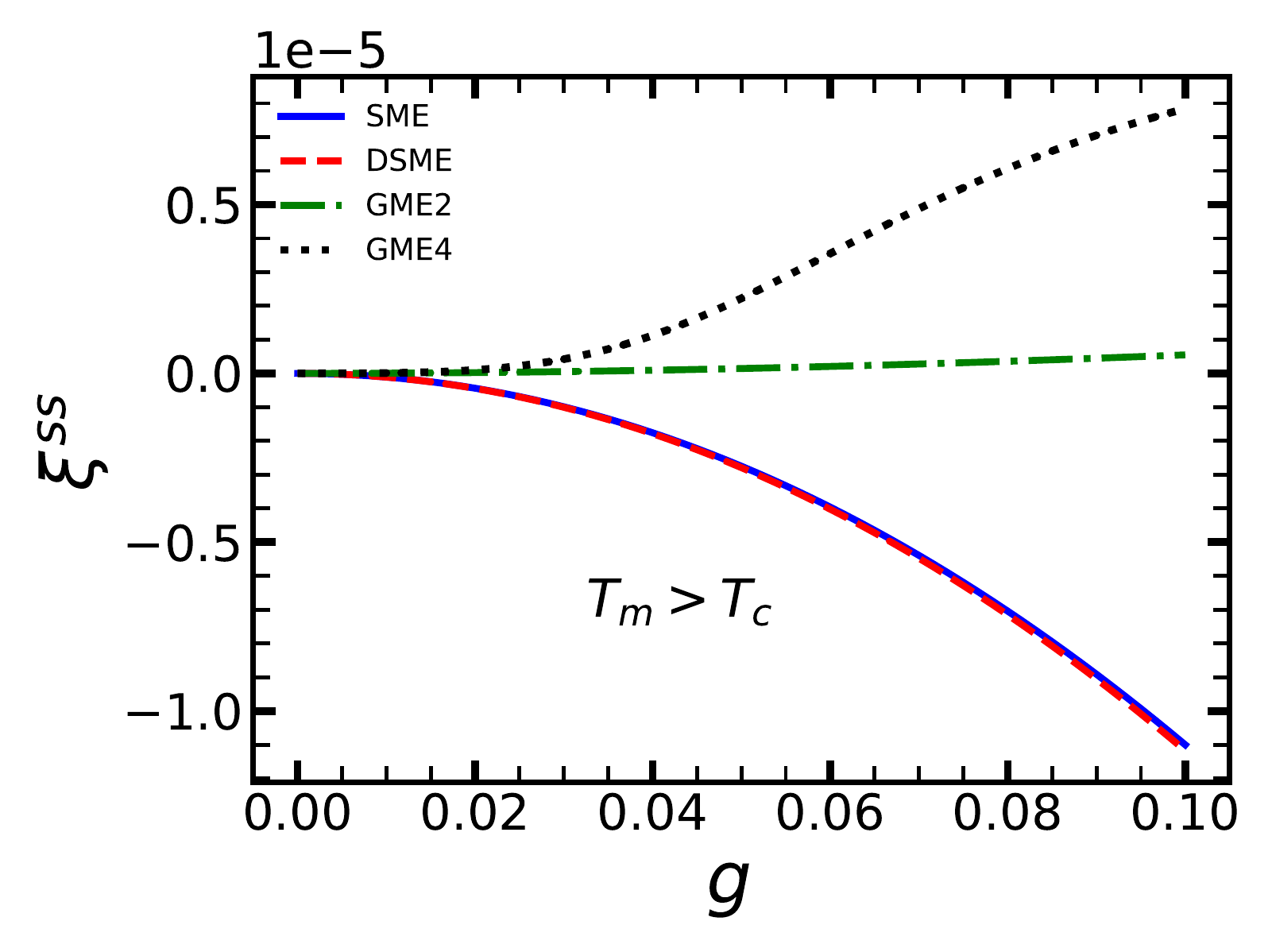}\label{fig:TmgtTcb}}	
  \caption{(a)~The steady state heat currents $\mathcal{J}^\text{ss}_\text{c}$ and $\mathcal{J}^\text{ss}_\text{m}$, and (b)~the rates of entropy production $\xi^\text{ss}$ as a function of normalized coupling strength $g$ for $T_\text{c} < T_\text{m}$. The results from the standard master equation (SME) are illustrated by the solid blue curves, those for the dressed-state master equation (DSME) by the red dashed curves, and those for the global master equation (GME) by the green dash--dotted curves (two side-bands) and the black dotted curves (four side-bands). Parameters: $T_\text{c}=101$\,mK and $T_\text{m}=106$\,mK, with the rest of the parameters as in Fig.\ \ref{fig:TmltTc}.}
  \label{fig:TmgtTc}
\end{figure}

\subsection{Equal temperatures, $T_\mathrm{c} = T_\mathrm{m}$}
When both baths are at the same temperature, the second law of thermodynamics dictates that the heat currents $\mathcal{J}_\text{c}$ and $\mathcal{J}_\text{m}$ must stay zero all the time. However, we find that the SME and the DSME predict non-zero values of $\mathcal{J}^\text{ss}_\text{c}$ and $\mathcal{J}^\text{ss}_\text{m}$, as shown in Fig.~\ref{fig:TmeqTc}; this represents a violation of the second law of thermodynamics as stated above. This sort of violation of second law when baths attached to system are kept at same temperature has been reported for Fermionic transport models~\cite{fermi1-violation, fermi2-violation} and two interacting harmonic oscillators~\cite{kosloff_local_violate}. We find that this unphysical result disappears if the system is described by the GME; when including the phonon side-bands terms in the master equation both heat currents $\mathcal{J}_\text{c}$ and $\mathcal{J}_\text{m}$ become zero. In Ref.~\cite{Patrick-LocalVsGlobal}, it was reported that when the baths are at same temperature
the global approach captures an accurate description of steady state of the system, whereas the local approach fails even in weak coupling. We expect, and indeed observe, similar results to hold in our case as well.

\subsection{Optics colder than mechanics, $T_\mathrm{c} < T_\mathrm{m}$}
For the case where $T_\text{c} < T_\text{m}$, Fig.~\ref{fig:TmgtTca} shows the steady state heat currents $\mathcal{J}^\text{ss}_\text{c}$ and $\mathcal{J}^\text{ss}_\text{m}$ as a function of the normalized coupling strength $g$. We see that, for all values of $g$, the two heat currents are equal and are therefore consistent with energy conservation. Moreover, for $g=0$ there is no heat flow into or out of the system, as required, and as $g$ increases the heat currents increase in magnitude. According to the Clausius statement of second law, heat must flow from the hot to the cold bath. Figure~\ref{fig:TmgtTca}, however, shows that if the dynamics of the system is described by the SME or the DSME, heat flows from the cold to the hot bath, independent of the optomechanical coupling strength. It is only when we include the phonon side-modes in the master equation that the direction of the heat current becomes correct and the violation of the second law of thermodynamics disappears.

The entropy production rate is plotted as a function of the coupling strength $g$ in Fig.~\ref{fig:TmgtTcb}. Since the entropy production must be non-negative, this figure yet again demonstrates that the local (SME) and the semi-global (DSME) approaches are inconsistent with second law of thermodynamics, even in the weak coupling regime. Inclusion of phonon side-bands, as per the GME, recovers consistency with the second law of thermodynamics in both weak and strong coupling regimes. Further investigation reveals that the source of this inconsistency is the assumption of an unphysical flat bath spectrum and the presence of non-secular terms, both of which are relaxed when deriving the GME. Indeed, if we reduce the number of side-bands or consider a flat response function for the baths, then even the dynamical description of the system by the GME results in a violation of the second law of thermodynamics.

\begin{figure}[t!]
  \centering
  \includegraphics[width=0.5\textwidth]{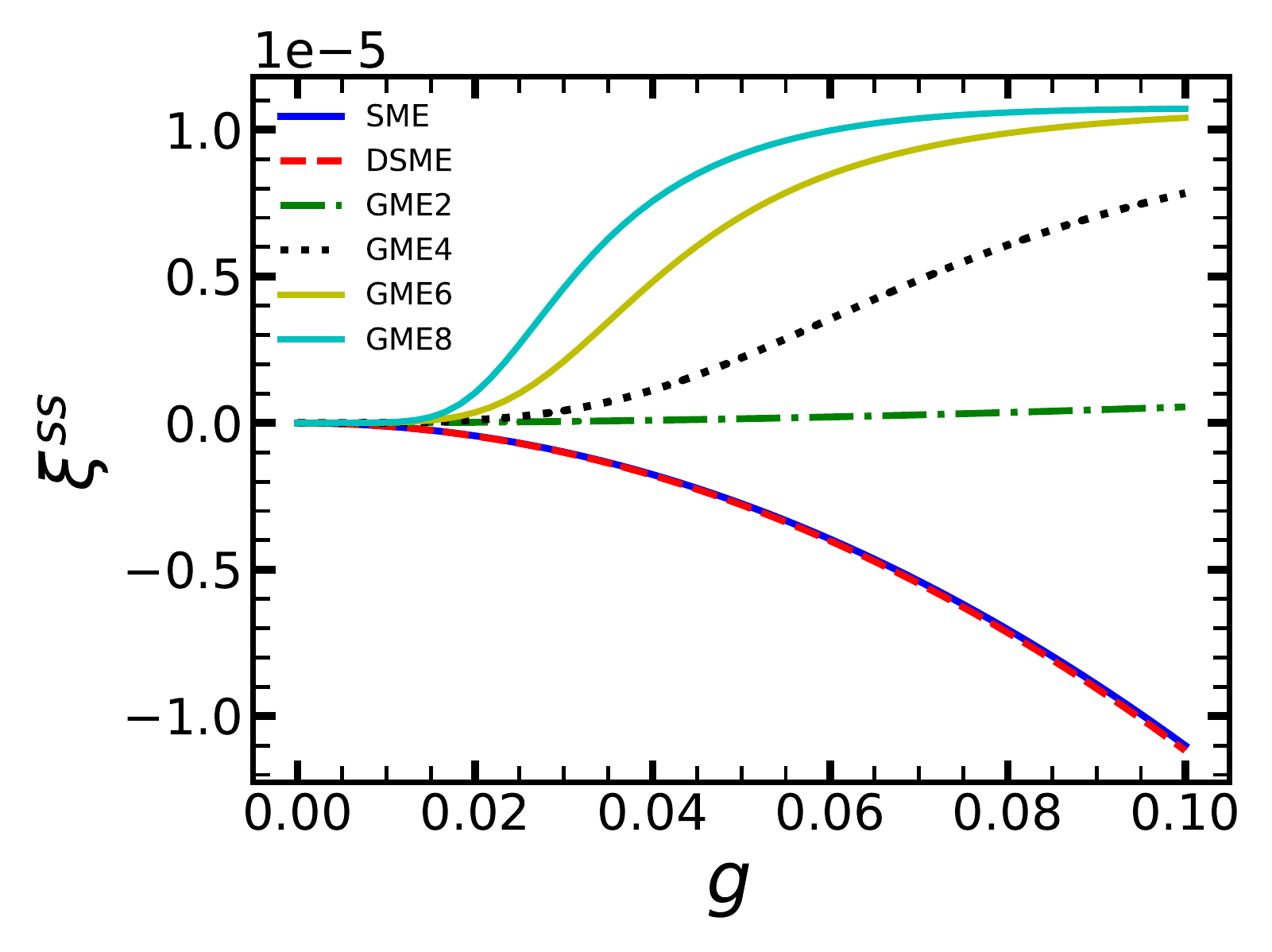}
  \caption{The rates of entropy production $\xi^\text{ss}$ as a function of normalized coupling strength $g$. From the lowest curve upward, the results shown are those for the dressed-state master equation (DSME), the standard master equation (DSME), and for the global master equation (GME$n$) with two ($n=2$), four ($n=4$), six ($n=6$), and eight ($n=8$) side-bands. All parameters are as in Fig.\ \ref{fig:TmgtTc}.}
  \label{fig:SB}
\end{figure}

To demonstrate the robustness of our investigation and the convergence of the GME as the number of sidebands is increased, we plot in Fig.~\ref{fig:SB} the rates of entropy production for a number of different situations. Convergence of the GME requires an increasing number of side-bands as $g$ increases, but the higher side-bands do not contribute significantly in the regime where $g\ll\omega_\text{m}$. In any case, the inclusion of more side-bands does not change the qualitative behavior of the heat currents or the steady-state entropy production.

\section{Conclusions}\label{sec:conclusion}
We have studied a generic optomechanical system coupled to two heat baths, one optical and one mechanical. Our focal point was the consistency of its dynamics with second law of thermodynamics, with respect to local and non-local approaches to obtaining the master equation describing this dynamics. When the two baths attached to the system are kept at same temperature, a non-zero heat current persists in steady state in the standard and dressed-state pictures, which violates the second law of thermodynamics. On the other hand, when the system is described by means of a non-local dynamical equation where each heat bath acquires a global character, the heat currents vanish. When the mechanical bath is held at a higher temperature, we have also seen that the steady-state heat current flows from the cold to the hot bath, and that there is a negative entropy production rate in steady state, under the first two descriptions. These violations are present in both the weak and strong coupling regimes that are typically investigated using these descriptions. In order to obtain a thermodynamically consistent and physically plausible result we accounted for the phonon side-bands in the master equation whilst taking into account the frequency-dependence of the bath occupation number. This corrects the direction of the heat flow and restores consistency with the second law of thermodynamics.

In closing, we note that the implications of our results are rather broad. Regardless of the regime in which one operates, thermodynamic quantities such as the rate of entropy production deduced from the standard or dressed-state master equations differ quantitatively, if not qualitatively, from those deduced from the global master equation. Due to the consistency of the latter with the second law of thermodynamics, we are drawn to the conclusion that when the mechanical bath is hotter than the optical bath, and especially outside the weak-coupling regime, any thermodynamic prediction for optomechanical systems should be based on the global master equation.

\begin{acknowledgements}
The authors are grateful to R.\ Kosloff, A.\ \.{I}mamo\u{g}lu, V.\ Vedral, C.\ Marletto, D.\ Hu, K.\ Brandner, and M.\ Paternostro for valuable discussions. M.\ T.\ N.\ and \"{O}.\ E.\ M.\ acknowledge support by the Scientific and Technological Research Council of Turkey (T\"{U}B{\.I}TAK), Grant No.\ 117F097), and by the COST Action Quantum Technologies with Ultra-Cold Atoms (CA16221). A.\ X.\ acknowledges funding from the European Union's Horizon 2020 research and innovation programme under Grant Agreement No.\ 732894 (FET Proactive HOT).
\end{acknowledgements}

\appendix

\section{Dynamics of the system using the SME or the DSME}\label{app:dynamics}
The equations of motions for the relevant dynamical observables of our system are determined from the DSME, Eq.~(\ref{eq:DSME}). The equations of motion read
\begin{align*}\label{eq:dynamics}
\tfrac{\rmd}{\rmd t}\langle\hat n_\text{c}\rangle&=\kappa_\text{c}(\bar{n}_\text{c}-\langle\hat n_\text{c}\rangle), \\
\tfrac{\rmd}{\rmd t}\langle\hat p\rangle&=-\omega_\text{m}\langle\hat q\rangle-\tfrac{\kappa_\text{m}}{2}\langle\hat p\rangle+2g\langle\hat n_\text{c}\rangle,\\
\tfrac{\rmd}{\rmd t}\langle\hat q\rangle&=\omega_\text{m}\langle\hat p\rangle-\tfrac{\kappa_\text{m}}{2}\langle\hat q\rangle +\kappa_\text{m}\alpha\langle\hat n_\text{c}\rangle,\\
\tfrac{\rmd}{\rmd t}\langle(\Delta\hat n_\text{c})^2\rangle&=\kappa_\text{c}\bar{n}_\text{c}+(2\kappa_\text{c}\bar{n}_\text{c}+\kappa_\text{c})
\langle\hat n_\text{c}\rangle-2\kappa_\text{c}\langle(\Delta\hat n_\text{c})^2\rangle, \\
\tfrac{\rmd}{\rmd t}\langle \hat n_\text{c},\hat p\rangle&=-\kappa\langle \hat n_\text{c},\hat p\rangle-\omega_\text{m}\langle
\hat n_\text{c},\hat q\rangle+2g\langle(\Delta\hat n_\text{c})^2\rangle,\\
\tfrac{\rmd}{\rmd t}\langle \hat n_\text{c},\hat q\rangle&=-\kappa\langle \hat n_\text{c},\hat q\rangle+\omega_\text{m}\langle
\hat n_\text{c},\hat p\rangle +\kappa_\text{m}\alpha\langle(\Delta\hat n_\text{c})^2\rangle,\\
\tfrac{\rmd}{\rmd t}\langle \hat n_\text{m}\rangle&=-\kappa_\text{m}(\langle \hat n_\text{m}\rangle-\bar n_\text{m})+
g(\langle \hat n_\text{c},\hat p\rangle+\langle \hat n_\text{c}\rangle\langle\hat p\rangle)\\
&\qquad+\tfrac{\kappa_\text{m}}{2}\alpha(\langle \hat n_\text{c},\hat q\rangle+\langle\hat n_\text{c}\rangle\langle\hat q\rangle),\\
\tfrac{\rmd}{\rmd t}\langle \hat n_\text{c}\hat p\rangle&=-\kappa\langle\hat{n}_\text{c}\hat{p}\rangle-\omega_\text{m}\langle\hat{n}_\text{c}\hat{q}\rangle+2g\langle\hat{n}^2_\text{c}\rangle+\kappa_\text{c}\bar{n}_\text{c}\langle\hat{p}\rangle,\\
\tfrac{\rmd}{\rmd t}\langle \hat n_\text{c}\hat q\rangle&=-\kappa\langle\hat{n}_\text{c} q\rangle+\alpha\kappa_\text{m}\langle\hat{n}^2_\text{c}\rangle+\kappa_\text{c}\bar{n}_\text{c}\langle q\rangle\\
&\qquad+\omega_\text{m}\langle\hat{n}_\text{c}\hat{p}\rangle,\ \text{and}\\
\tfrac{\rmd}{\rmd t}\langle\hat{n}^2_\text{c}\rangle&= \kappa_\text{c}\bar{n}_\text{c}-2\kappa_\text{c}\langle\hat{n}^2_\text{c}\rangle+\kappa_\text{c}(4\bar{n}_\text{c}+1)\langle\hat{n}_\text{c}\rangle,
\end{align*}
where $\alpha=g/\omega_\text{m}$, $\hat{q}=\hat{b}+\hat{b}^\dag$, $\hat{p}=i(\hat{b}-\hat{b}^\dag)$, and $\kappa=\kappa_\text{c}+\kappa_\text{m}/2$. The correlation functions between any two operators $\hat{o}_1$ and $\hat{o}_2$ are denoted by $\langle \hat{o}_1,\hat{o}_2\rangle:=\langle\hat{o}_1\hat{o}_2\rangle-\langle \hat{o}_1\rangle\langle\hat{o}_2\rangle$. The steady-state solutions of these dynamical equations in the long time limit are as follows:
\begin{align*}
\langle\hat n_\text{c}\rangle^\text{ss}&=\bar n_\text{c},\\
\langle\hat q\rangle^\text{ss} &=\biggl(\frac{8g\omega_\text{m}+2\alpha\kappa_\text{m}^2}{4\omega_\text{m}^2+\kappa_\text{m}^2}\biggr)\langle\hat n_\text{c}\rangle^\text{ss},\\
\langle\hat p\rangle^\text{ss} &=\biggl(\frac{4g\kappa_\text{m}-4\alpha\omega_\text{m}\kappa_\text{m}}{4\omega_\text{m}^2+\kappa_\text{m}^2}\biggr)\langle\hat n_\text{c}\rangle^\text{ss},\\
\langle(\Delta\hat n_\text{c})^2\rangle^\text{ss}&=\bar{n}_\text{c}(\bar{n}_\text{c}+1),\\
\langle \hat n_\text{c},\hat p\rangle^\text{ss}&=\frac{2g\kappa_\text{c}}{\omega_\text{m}^2+\kappa^2}\langle(\Delta\hat n_\text{c})^2\rangle^\text{ss},\\
\langle \hat n_\text{c},\hat q\rangle^\text{ss}&=\biggl(\frac{2g\omega_\text{m}+\alpha\kappa_\text{m}\kappa}{\omega_\text{m}^2+\kappa^2}\biggr)\langle(\Delta\hat n_\text{c})^2\rangle^\text{ss},\\
\langle \hat n_\text{m}\rangle^\text{ss}&=\bar n_\text{m}
+\frac{g}{\kappa_\text{m}}(\langle \hat n_\text{c},\hat p\rangle^\text{ss}+\langle\hat n_\text{c}\rangle^\text{ss}\langle\hat p\rangle^\text{ss})\\
&\qquad+\frac{\alpha}{2}(\langle \hat n_\text{c},\hat q\rangle^\text{ss}+\langle\hat n_\text{c}\rangle^\text{ss}\langle\hat q\rangle^\text{ss}),\\
\langle \hat n_\text{c}\hat q\rangle^\text{ss}&=A\biggl[\kappa_\text{c}\bar{n}_\text{c}(\kappa\langle q\rangle^\text{ss}+\omega_\text{m}\langle p\rangle^\text{ss})\\
&\qquad\quad+(\kappa\alpha\kappa_\text{m}+2 g\omega_\text{m})
  \langle\hat{n}^2_\text{c}\rangle^\text{ss}\biggr],\\
\langle \hat n_\text{c}\hat p\rangle^\text{ss}&=A\biggl[\kappa_\text{c}\bar{n}_\text{c}(\kappa\langle p \rangle^\text{ss}-\omega_\text{m}\langle q\rangle^\text{ss})\\
&\qquad\quad+(2\kappa g-\alpha\kappa_\text{m}\omega_\text{m})\langle\hat{n}^2_\text{c}\rangle\biggr],\ \text{and}\\
\langle \hat{n}^2_\text{c}\rangle^\text{ss}&=\bar{n}_\text{c}(2\bar{n}_\text{c}+1),
\end{align*}
where $A=1\big/\bigl(\kappa^2+\omega^2_\text{m}\bigr)$ and $\bar{n}_\text{c}:=\bar{n}_\text{c}(\omega_\text{c})$. The corresponding equations for the SME are found by setting $\alpha=0$.

\section{Heat currents using the SME or the DSME}\label{app:heatflow}
The heat currents to the optical and mechanical baths to which the optomechanical system is attached are, respectively,
\begin{align*}
\mathcal{J}_\text{c}&=\kappa_\text{c}(\omega_\text{c}-g\langle\hat{q}\rangle)(\bar{n}_\text{c}-\langle\hat{n}_\text{c}\rangle) + g\kappa_\text{c}\langle\hat{n}_\text{c},\hat{q}\rangle,\ \text{and}\\
\mathcal{J}_\text{m}&=\omega_\text{m}\kappa_\text{m}(\bar{n}_\text{m}-\langle\hat{n}_\text{m}\rangle)+g\kappa_\text{m}(\langle\hat{n}_\text{c},\hat{q}\rangle + \langle\hat{n}_\text{c}\rangle\langle\hat q\rangle)\\
&\qquad-g\alpha\kappa_\text{m}(\langle(\Delta\hat{n}_\text{c})^2\rangle+{\langle\hat{n}_\text{c}\rangle}^2).
\end{align*}
At steady state the heat currents are therefore given by
\begin{align*}
\mathcal{J}_\text{c}^\text{ss}&=g\kappa_\text{c}\langle\hat{n}_\text{c},\hat{q}\rangle^\text{ss}\ \text{and}\\
\mathcal{J}_\text{m}^\text{ss}&=-g\kappa_\text{c}\langle\hat{n}_\text{c},\hat{q}\rangle^\text{ss}.
\end{align*}
The results for the SME are once again obtained by setting $\alpha=0$.

\end{document}